\documentclass[twocolumn,showpacs,amsmath,amstex,amssymb,mathfonts,prb]{revtex4}
\usepackage{graphicx}

\newcommand{\brr}{{\bf r}}

\begin{document}

\title{Theory of tunneling transport in periodic chains}

\author{Emil Prodan$^1$ and Roberto Car$^2$}
\address{$^1$Department of Physics, Yeshiva University, New York, NY 10016} 
\address{$^2$Department of Chemistry and Princeton Institute fot the Science and Technology of Materials, Princeton University, Princeton, NJ 08544}

\begin{abstract}
We present an extended discussion of a recently proposed theoretical approach for off-resonance tunneling transport. The proofs and the arguments are explained at length and simple analogies and illustrations are used where possible. The result is an analytic formula for the asymptotic tunneling conductance which involves the overlap of three well defined physical quantities. We argue that the formula can be used to gain fresh insight into the tunneling transport characteristics of various systems. The formalism is applied here to molecular devices consisting of planar phenyl chains connected to gold electrodes via amine linkers.  
\end{abstract}

\date{\today}

\maketitle

Exciting new developments in molecular transport have lead to accurate single molecule measurements. A large part of the experimental data has been confirmed by several independent experimental groups.\cite{Venkataraman:2006db,Venkataraman:2006lq,Hybertsen:2008oj,Chen:2006fk,Chu:2007uk,Kiguchi:2008eq} We are particularly interested in the data for molecules made of several repeating units or monomers, such as alkyl, phenyl, acene chains of various lengths. For such systems, we develop a semi-analytic theory of tunneling transport.

The signature of the tunneling transport is exponential dependence, $G$=$G_ce^{-\beta N}$, of the conductance $G$ on the number of monomers $N$. In the past, the off-resonant tunneling transport was described and understood in terms of effective electrons tunneling through square barriers.\cite{Simmons:1963fh} Such treatment works well as long as the effective mass approximation remains valid at the Fermi level. However, many systems, in particular organic chains, display large insulating gaps and flat bands and very often the effective mass approximation for these systems fails when one moves away from the band edges. The modern theory of tunneling transport \cite{Mavropoulos:2000cr,Tomfohr:2002oq,Tomfohr:2004ve,Fagas:2004uq} connects the tunneling exponent $\beta$ to the complex band structure of the chains, an approach that goes well beyond the effective mass treatments. In a past publication,\cite{Prodan:2007qv} we have contributed to the picture by deriving an expression for $G_c$, the contact conductance.

In this work, we extend our previous discussion of the off-resonant tunneling transport in periodic insulating chains. We carefully review our previous arguments, extend them when necessary and simplify them when possible. In addition, we give a more detailed discussion of the conductance within the Time Dependent Current-Density Functional Theory (TDCDFT) for which we present a formally exact result.

The theory is applied to molecular devices consisting of planar phenyl chains linked to gold wires via amine anchoring groups. We report theoretical values of the linear conductance for devices containing up to 4 phenyl rings, which are compared with the experimental data of Ref.~\onlinecite{Venkataraman:2006db}. Based on our analytic expression for $G_c$, we discuss and quantify the main factors influencing the charge transport in these devices. We recall that a similar study was recently completed for molecular devices involving amine linked alkyl chains.\cite{Prodan:2008by}

\section{Transport: General considerations}

We consider a charge transport experiment involving a device made of a molecular chain attached to metallic leads (see Fig.~\ref{setup}). The system is driven by a small time oscillating electric field ${\bf E}_1^{\mbox{\tiny{ext}}}(r,t)$, whose effects are treated in the linear response regime. The dc regime is obtained by letting the frequency of the oscillation go to zero. The existence of a steady state is implicitly assumed.

Within the Time Dependent Current-Density Functional Theory (TDCDFT) and linear response regime, the current density is given by:\cite{G.-Vignale:1996fk,Vignale:1997fk}
\begin{equation}\label{dcurrent}
{\bf j}(\brr,\omega)=\int \hat{\sigma}^{\mbox{\tiny{KS}}}(\brr,\brr';\omega) {\bf E}^{\mbox{\tiny{eff}}}_1(\brr',\omega)d\brr',
\end{equation}
where $\hat{\sigma}^{\mbox{\tiny{KS}}}$ is the equilibrium Kohn-Sham conductivity tensor. A local density approximation expression for ${\bf E}^{\mbox{\tiny{eff}}}_1(\brr,\omega)$ is given in Ref.~\onlinecite{Vignale:1997fk}:
\begin{equation}\label{Eeff}
{\bf E}^{\mbox{\tiny{eff}}}_1= \frac{1}{e}\nabla \phi_1^{\mbox{\tiny{ext}}}+\frac{1}{e}\nabla \phi^{\mbox{\tiny{HXC}}}_1+{\bf E}^{\mbox{\tiny{dyn}}}_1,
\end{equation}
where $\phi^{\mbox{\tiny{HXC}}}_1$ is the linearized Hartree-exchange-correlation potential of the equilibrium DFT and ${\bf E}^{\mbox{\tiny{dyn}}}_1$ is the dynamical part of ${\bf E}^{\mbox{\tiny{eff}}}_1$, given by ${\bf E}^{\mbox{\tiny{dyn}}}_1$=$-\frac{1}{en_0}\nabla \hat{\zeta}$, with $\hat{\zeta}$ the viscoelastic stress tensor. In the linear regime:
\begin{equation}
\begin{array}{c}
\int d\brr' \hat{\sigma}^{\mbox{\tiny{KS}}}(\brr,\brr') {\bf E}^{\mbox{\tiny{dyn}}}_1(\brr') \medskip \\ 
=\int d\brr' \int d\brr'' \hat{\sigma}^{\mbox{\tiny{KS}}}(\brr,\brr') \hat{{\cal F}}(\brr',\brr''){\bf j}(\brr''),
\end{array}
\end{equation}
where
\begin{equation}
{\cal F}_{\alpha \beta}(\brr,\brr') \equiv  \left . \frac{\delta E_\alpha^{\mbox{\tiny{dyn}}}(\brr)}{\delta j_\beta (\brr')}\right |_{\phi_1^{\mbox{\tiny{ext}}}=0}.
\end{equation}
$\hat{{\cal F}}(\brr,\brr')$ is understood as a matrix with elements ${\cal F}_{\alpha \beta}(\brr,\brr')$ and matrix multiplication is understood between $\hat{\sigma}$ and $\hat{\cal F}$ and between $\hat{\cal F}$ and ${\bf j}$. This leads to
 \begin{equation}
 \begin{array}{c}
 {\bf j}(\brr) = \int d \brr' \int d \brr'' [1- \hat{\sigma}^{\mbox{\tiny{KS}}}*\hat{{\cal F}}]^{-1}(\brr,\brr') \medskip \\
\times  \hat{ \sigma}^{\mbox{\tiny{KS}}}(\brr',\brr'') \nabla \phi^{\mbox{\tiny{ad}}}(\brr''),
\end{array}
 \end{equation}
which is an RPA type expression for the current density.  Here
\begin{equation}
\phi^{\mbox{\tiny{ad}}}(\brr'')=\phi_1^{\mbox{\tiny{ext}}}+\phi^{\mbox{\tiny{HXC}}}_1
\end{equation}
is the driving potential plus the adiabatic response of the electrons.

The net current flowing through the device is given by
\begin{equation}\label{netcurrent}
\begin{array}{c}
I=\int_{\Sigma} d{\bf S}  \ \int d \brr' \int d \brr'' [1- \hat{\sigma}^{\mbox{\tiny{KS}}}*\hat{{\cal F}}]^{-1}(\brr,\brr') \medskip \\
\times \hat{\sigma}^{\mbox{\tiny{KS}}}(\brr',\brr'') \nabla \phi^{\mbox{\tiny{ad}}}(\brr''),
\end{array}
\end{equation}
where $\Sigma$ is an arbitrary transversal section. The potential drop that is measured by a voltmeter attached to the two ends of the device is given by:\cite{A.-Kamenev:2001uq}
\begin{equation}\label{voltmeter}
\Delta \phi = [\phi_{\mbox{\tiny{ext}}}+\phi_1^{\mbox{\tiny{H}}}]_{+\infty}-[\phi_{\mbox{\tiny{ext}}}+\phi_1^{\mbox{\tiny{H}}}]_{-\infty},
\end{equation}
and the linear conductance is defined as $G=\frac{I}{\Delta \phi}$. Note that the screening also contributes to the potential drop. Here, $\phi_1^{\mbox{\tiny{H}}}$ is the Hartree potential corresponding to the density perturbation $n_1$. 

\subsection{An exact expression for linear conductance}

We now show that $\Delta \phi$ can be pulled out of the complicated integrals in Eq.~\ref{netcurrent}. For this, let us restrict the integral over $d \brr''$ in Eq.~\ref{netcurrent} to a volume between two distant sections $\Sigma_-$ and $\Sigma_+$. We will later take these surfaces to infinity. Now, because
\begin{equation}\label{fproperty}
\sum_\alpha \partial_\alpha \sigma_{\alpha \beta}^{\mbox{\tiny{KS}}}(\brr,\brr') =\sum_\beta \partial'_\beta \sigma_{\alpha \beta}^{\mbox{\tiny{KS}}}(\brr,\brr')=0,
\end{equation}
we have
\begin{equation}
\hat{\sigma}^{\mbox{\tiny{KS}}}(\brr',\brr'')\nabla''\phi^{\mbox{\tiny{ad}}}(\brr'')=\nabla'' \hat{\sigma}^{\mbox{\tiny{KS}}}(\brr',\brr'')\phi^{\mbox{\tiny{ad}}}(\brr''),
\end{equation}
and we can transform the integral over $\brr''$ in Eq.~\ref{netcurrent} in a surface integral:
\begin{equation}
\begin{array}{c}
 I= \int_{\Sigma} d{\bf S} \int d \brr'  ( \int\limits_{\Sigma_+}d{\bf S}''-\int\limits_{\Sigma_-}d{\bf S}'' )  \medskip \\
 \times  [1- \hat{\sigma}^{\mbox{\tiny{KS}}}*\hat{{\cal F}}]^{-1}(\brr,\brr') \hat{\sigma}^{\mbox{\tiny{KS}}}(\brr',\brr'') \phi^{\mbox{\tiny{ad}}}(\brr'') .
\end{array}
 \end{equation}
We now chose the sections $\Sigma_\pm$ to be iso-surfaces of $\phi^{\mbox{\tiny{ad}}}$ in which case:
\begin{equation}
\begin{array}{c}
 I = \int_{\Sigma} d{\bf S}\int d \brr'  ( \phi^{\mbox{\tiny{ad}}}_+\int\limits_{\Sigma_+}d{\bf S}''-\phi^{\mbox{\tiny{ad}}}_-\int\limits_{\Sigma_-} d{\bf S}'')  \medskip \\
\times  [1- \hat{\sigma}^{\mbox{\tiny{KS}}}*\hat{{\cal F}}]^{-1}(\brr,\brr') \hat{\sigma}^{\mbox{\tiny{KS}}}(\brr',\brr'').
 \end{array}
 \end{equation}
 But once we pulled the potential out, the surface integrals no longer depend on the shape and position of the surfaces, a fact that follows from the property in Eq.~\ref{fproperty}. Therefore, we can deform $\Sigma_\pm$ into one single surface $\Sigma''$ to obtain:
\begin{equation}
\begin{array}{c}
 I =\Delta \phi^{\mbox{\tiny{ad}}} \int_{\Sigma} d{\bf S} \int d \brr'  \int_{\Sigma''} d {\bf S}'' \medskip \\
 \times [1- \hat{\sigma}^{\mbox{\tiny{KS}}}*\hat{{\cal F}}]^{-1}(\brr,\brr') \hat{\sigma}^{\mbox{\tiny{KS}}}(\brr',\brr'').
 \end{array}
 \end{equation}
At this point, let us write the explicit expression of $\phi^{\mbox{\tiny{ad}}}$:
\begin{equation}\label{deltaf}
\begin{array}{c}
\phi^{\mbox{\tiny{ad}}}(\brr)=\phi_1^{\mbox{\tiny{ext}}}(\brr)+\int d\brr' \frac{n_1(\brr')}{|\brr-\brr'|} \medskip \\
+\int d \brr' \left . \frac{\delta v^{\mbox{\tiny{XC}}}(\brr)}{\delta n(\brr')}\right |_{\phi_1^{\mbox{\tiny{ext}}}=0} n_1(\brr').
\end{array}
\end{equation}
The density $n_1$ is localized near the junction, but its decay away from the junction can be rather slow. Due to the long range of the Coulomb kernel, the Hartree potential will take finite values at $\pm \infty$ and will contribute to $\Delta \phi^{\mbox{\tiny{ad}}}$. The contribution from xc part was discussed in Refs.~\onlinecite{Koentopp:2006uo} and \onlinecite{Koentopp:2006ys}. Here, it was pointed out that the common density functionals use semi-local exchange-correlation potentials in which case the kernel $\delta v_{\mbox{\tiny{xc}}}(\brr)/\delta n(\brr')$ decays extremely fast with the separation $|\brr-\brr'|$ and therefore the last integral in Eq.~\ref{deltaf} will vanish when $\brr$ is taken at $\pm \infty$. The conclusion is that $\Delta \phi^{\mbox{\tiny{ad}}}$ is in fact the potential drop measured by a voltmeter (see Eq.~\ref{voltmeter}).

However, in same references it was also pointed out that functionals like those involving exact exchange lead to kernels  $\delta v^{\mbox{\tiny{XC}}}(\brr)/\delta n(\brr')$ slowly decaying with $|\brr-\brr'|$, in which case the last integral in Eq.~\ref{deltaf} will take finite values when $\brr$ is taken at $\pm \infty$. In this case, we have to treat $\phi^{\mbox{\tiny{XC}}}_1$ as we treated ${\bf E}^{\mbox{\tiny{dyn}}}_1$, in which case ${\cal F}_{\alpha \beta}(\brr,\brr') $ has to be redefined as:
 \begin{equation}
{\cal F}_{\alpha \beta}(\brr,\brr') \equiv  \left . \frac{ \delta [E_\alpha^{\mbox{\tiny{dyn}}}+ \partial_\alpha \phi_1^{\mbox{\tiny{XC}}}](\brr)}{\delta j_\beta (\brr')}\right |_{\phi_1^{\mbox{\tiny{ext}}}=0}
\end{equation} 
This expression has to be computed at finite frequencies first, where one will use the relation $n_1=\frac{1}{i\omega}\nabla {\bf j}$, and then the limit $\omega$$\rightarrow$0 has to be considered. Of course, in this case one can no longer use the local approximation of ${\bf E}^{\text{\tiny{dyn}}}$ given in Ref.~\onlinecite{Vignale:1997fk}.

In either case, we arrive at the following formally exact expression of the linear conductance:
 \begin{equation}\label{exactg}
 G = \int d \brr_\bot   \int d \brr'_\bot  \  [(1- \hat{\sigma}^{\mbox{\tiny{KS}}}*\hat{{\cal F}})^{-1}* \hat{\sigma}^{\mbox{\tiny{KS}}}]_{zz}(\brr_\bot,z;\brr'_\bot,z').
 \end{equation}
 Here, $\brr_\bot$ and $\brr'_\bot$ denote the coordinates of two normal surfaces to the axis of the device. The position of these two surfaces can be taken arbitrarily.
 
 \subsection{Linear conductance in adiabatic approximation}
 
 The adiabatic approximation neglects the dynamical effects, which is equivalent to setting $\hat{{\cal F}}$ to zero. In this case, the expression for the linear conductance becomes:
 \begin{equation}\label{adiabaticg}
 G = \int d \brr_\bot   \int d \brr'_\bot  \  \hat{\sigma}^{\mbox{\tiny{KS}}}_{zz}(\brr_\bot,z;\brr'_\bot,z').
 \end{equation}
We should point out that this expression also assumes a rapidly decaying kernel $\delta v_{\mbox{\tiny{xc}}}(\brr)/\delta n(\brr')$ with the separation $|\brr - \brr'|$.  It is remarkable that, after the inclusion of electronic screening in $\Delta \phi$, the expression for $G$ remains formally identical to the one derived by Baranger and Stone\cite{Baranger:1989bs} for non-interacting electrons.
 
 In the rest of the paper, the conductance will be evaluated using Eq.~\ref{adiabaticg}, a choice that is largely dictated by practical considerations. It amounts to implicitly assume that the dynamical effects are small, an hypothesis that we are unable to support with rigorous arguments. A previous numerical study found that dynamical effects play only a minor role,\cite{Sai:2005uq} but this study used only the local density approximation for ${\bf E}^{\text{\tiny{dyn}}}$ and considered small junctions, while here we focus on long molecular chains. We also leave the issue of non-locality of the xc potential to future investigations. Although very interesting, these studies would be extremely challenging, particularly in the case of the large systems that are considered in this paper.
  
For the nonlocal $zz$ component of the conductivity tensor, we can work with the following expression:\cite{Fisher:1981kx,Baranger:1989bs}
\begin{equation}\label{basicsigma}
\sigma_{zz}(\brr,\brr')=-\frac{e^2\hbar^3}{8\pi m^2} 
\Delta G_{\epsilon_F}(\brr,\brr') \ \overleftrightarrow{\partial'_z} \ 
 \ \Delta G_{\epsilon_F}(\brr',\brr),
\end{equation}
where 
\begin{equation}\label{delta}
\Delta G_{\epsilon_F}(\brr,\brr')=G_{\epsilon_F+i\delta}(\brr,\brr')-\ G_{\epsilon_F-i \delta}(\brr,\brr'),
\end{equation} 
and $G_\epsilon$ is the Green's function $G_\epsilon$=$(\epsilon$-$H)^{-1}$ of the Kohn-Sham Hamiltonian describing the equilibrium of the entire device: $H$=$-\frac{\hbar^2}{2m}\nabla^2$+$V_{\mbox{\tiny{eff}}}$, $V_{\mbox{\tiny{eff}}}=v_{\mbox{\tiny{ps}}}+v^{\mbox{\tiny{HXC}}}[n]$ with $v_{\mbox{\tiny{ps}}}$ being the ions' pseudo-potential and $n$ the electron density at equilibrium. As pointed out in Refs.~\onlinecite{Fisher:1981kx} and \onlinecite{Baranger:1989bs}, $\sigma_{zz}$ contains additional terms but they cancel out after the integrations in Eq.~\ref{adiabaticg} and therefore can be neglected.

\section{An analytic expression for the tunneling conductance} 

\begin{figure}
\center
\includegraphics[width=8cm]{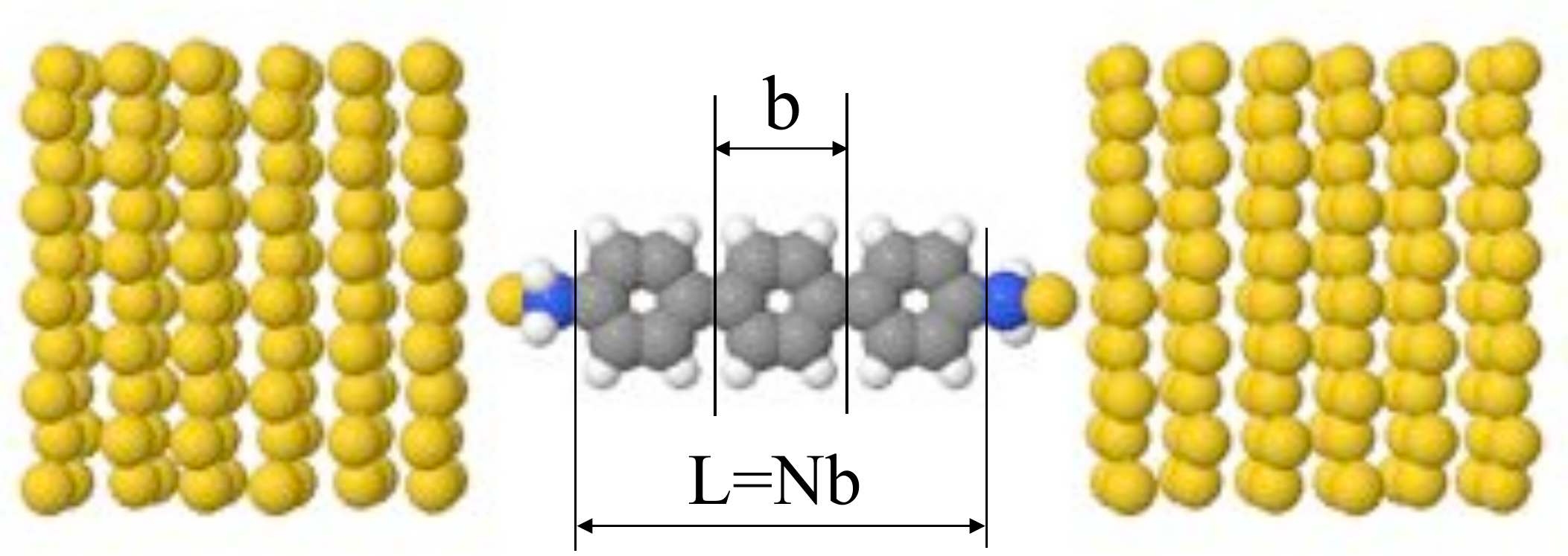}
\caption{Illustration of a typical device considered in this paper. The figure indicate the unit cell that is repeated periodically to obtain the periodic potential $V_0$. It also defines the length $L$ of the chain.}
 \label{setup}
\end{figure}

Consider a molecular device consisting of a long but finite periodic molecular chain (of unit cell $b$) attached to infinite metallic electrodes, like in Fig.~\ref{setup}. The orientation of the chain is along the $z$ axis. We assume that a self-consistent Kohn-Sham calculation has been completed for the entire device. The effective potential of the entire molecular device $V_{\mbox{\tiny{eff}}}$ is decomposed into a perfectly periodic piece, $V_0$, extending from $-\infty$ to $+\infty$, and a difference $\Delta V$=$V_{\mbox{\tiny{eff}}}-V_0$.  The periodic potential $V_0$ is constructed by periodically repeating the effective potential between $-b/2$ and $b/2$ at the middle of the chain. Our main assumption is that the potential difference $\Delta V$=$V_{\mbox{\tiny{eff}}}-V_0$ rapidly decays to zero inside the periodic chain. In other words, we assume that, to a very high degree, the effective potential inside the chain is periodic. This assumption proved to be accurate for the systems we studied so far, including the phenyl chains studied in this paper.

We regard the self-consistent Kohn-Sham Hamiltonian of the chain+leads as a periodic Hamiltonian,
 \begin{equation}\label{hzero}
 H_0=-\frac{\hbar^2}{2m}\nabla^2+V_0(\brr), \ V_0(\brr + b{\bf e}_z)=V_0(\brr),
 \end{equation}
strongly perturbed by the potential $\Delta V$. The effective Hamiltonian of the entire system is then
 \begin{equation}\label{hamil}
 H=H_0+\Delta V_L(\brr)+\Delta V_R(\brr),
 \end{equation}
where we divided  $\Delta V$ into left and right parts. We assume  that $\Delta V_{L,R}$ decay fast to zero as we move away from the contacts. We demonstrate in the following that, based on an analytic expression for the Green's function corresponding to $H_0$, we can derive an analytic, non-perturbative expression for the Green's function of the entire device. This is somewhat complementary to the approach presented in Ref.~\onlinecite{Smogunov:2004mp}, which views the devices as periodic leads perturbed by the junctions.

\subsection{Computing the Green's function for the periodic potential}

Let us first consider the Green's function $G_\epsilon^0$=$(\epsilon-H_0)^{-1}$, with $\epsilon$ outside the spectrum of $H_0$. To make the discussion more transparent, we recall that in 1 dimension, the Green's function for a Hamiltonian of the form $-\frac{\hbar^2}{2m}\frac{d^2}{dx^2} + V(x)$ can be conveniently written as:
\begin{equation}
G_\epsilon(x,x')=-\frac{2m}{\hbar^2}\frac{\psi_{<}(x_<) \psi_>(x_>)}{W(\psi_<,\psi_>)},
\end{equation}
with $x_<=\min(x,x')$ and $x_>=\max(x,x')$ and $W(\psi,\phi)$=$\psi \phi'$-$\phi\psi '$. Here, $\psi_<(x)$ and  $\psi_>(x)$ are the solutions of the Schrodinger equation:
\begin{equation}\label{1dgg}
[-\frac{\hbar^2}{2m}\frac{d^2}{dx^2}+V(x)]\psi(x) = \epsilon \psi(x)
\end{equation}
decaying to zero as $x\rightarrow -\infty$ and $x\rightarrow+\infty$, respectively. For a periodic system, the above expression reduces to:
\begin{equation}\label{1dg}
G_\epsilon(x,x')=-\frac{2m}{\hbar^2}\frac{\psi_{-k}(x_<) \psi_{k}(x_>)}{W(\psi_{-k},\psi_k)}
\end{equation}
where $\psi_k(x)$ is the Bloch function evaluated at the unique complex $k$ with Im[$k$]$>$0 for which the complex band energy satisfies $\epsilon_k=\epsilon$. To understand the simplicity of the above expressions, one should compare them with the formal expansion:
\begin{equation}\label{formal}
G_\epsilon(x,x')=\sum_n \frac{\phi_n(x)^*\phi_n(x')}{\epsilon_n -\epsilon}
\end{equation}
where \{$\phi_n(x)$,$\epsilon_n$\} is the infinite sequence of eigenvectors and corresponding eigenvalues of the Hamiltonian. As opposed to Eq.~\ref{1dgg}, in Eq.~\ref{formal} one has to compute a large number of wavefunctions and a truncation to $n$=$N$ will generate $O(N)$ errors. The expression shown in Eq.~\ref{1dgg} is generally valid only in 1 dimension. We will show in the following, however, that for periodic Hamiltonians we can derive this expression using the Riemann structure of the bands. Since the molecular chains in 3 dimensions still exhibit a Riemann structure,\cite{Prodan:2006yq} such derivation allows us to generalize Eq.~\ref{1dg} from strictly 1 dimension to molecular chains in 3 dimensions.

\begin{figure}
\center
\includegraphics[width=8.6cm]{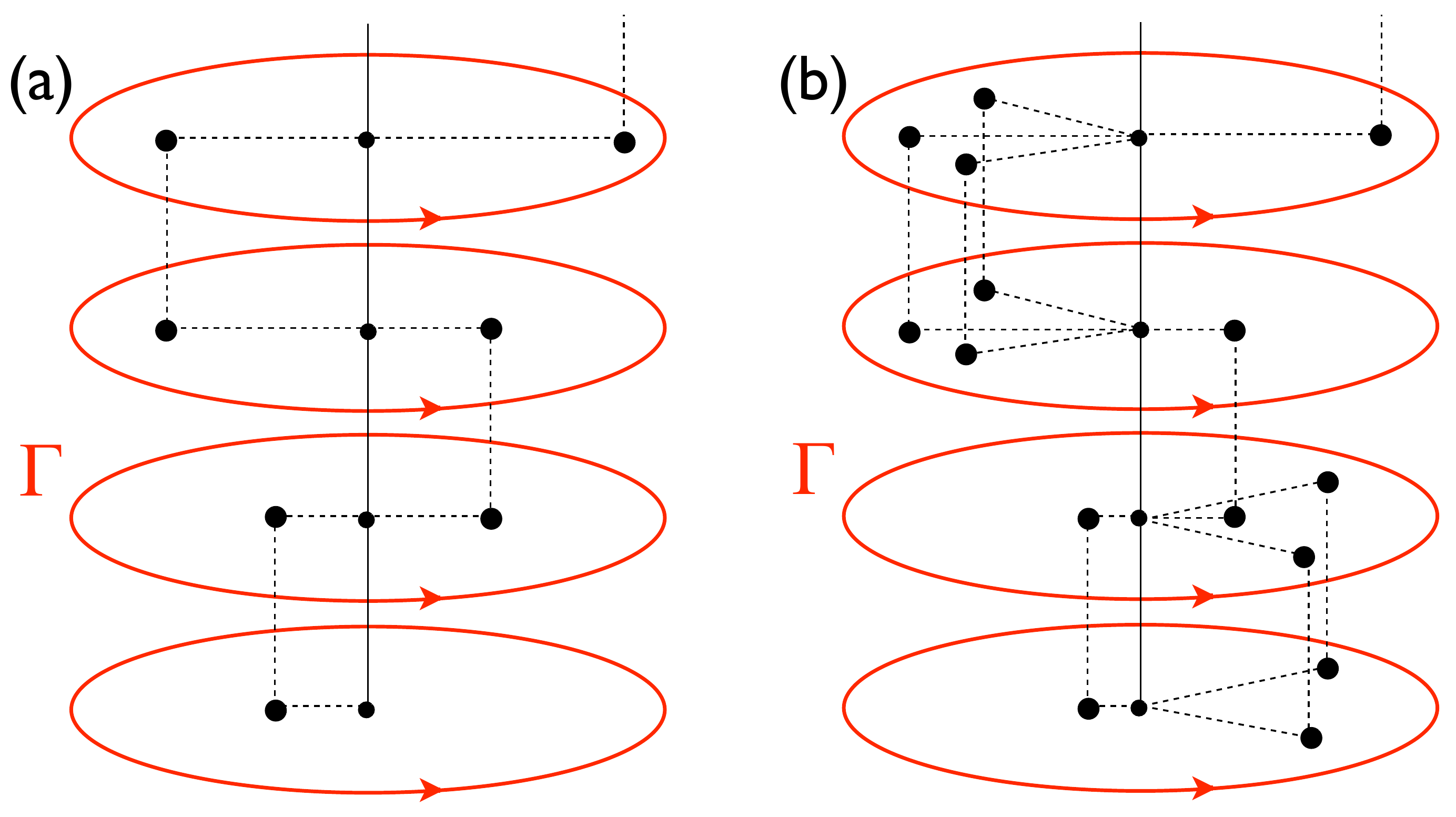}\\
\caption{The generic shape of the Riemann surface of the bands for strictly one dimensional periodic systems (left) and for periodic chains in 3 dimensions (right). The figure also illustrates the contour $\Gamma$ used in the main text.}
 \label{RiemannS}
\end{figure}

We start now the derivation. From Ref.~\onlinecite{Kohn:1959fk}, it is known that the Bloch function $\psi_\lambda(x)$ and the band energy $\epsilon_\lambda$ [$\lambda=\exp(ikb)$] can be defined on a Riemann surface that looks like in  Fig.~\ref{RiemannS}. This Riemann surface is made of a sequence of unit disks that are cut and then re-glued together as explained in Ref.~\onlinecite{Kohn:1959fk}. Different disks correspond to different bands and the physical, real $k$ bands can be generated by evaluating $\epsilon_\lambda$ along the unit circles of each disks. Starting from the eigenvalue-eigenvector expansion, where we use the standard normalization of the Bloch functions:
\begin{equation}
\frac{1}{b}\int_{-b/2}^{b/2} \psi_{n,-k}(x)\psi_{n,k}(x)dx=1,
\end{equation}
we can write:
\begin{equation}
G_\epsilon(x,x')= \frac{1}{2\pi}\sum_n \int dk \frac{\psi_{n,-k}(x)\psi_{n,k}(x')}{\epsilon-\epsilon_{n,k} } .
\end{equation}
By using the Riemann structure, we can combine the sum over the band index and the integration over $k$ into one single integral over a contour $\Gamma$ defined on the Riemann surface of the bands (see Fig.~\ref{RiemannS}a):
\begin{equation}
G_\epsilon(x,x')=\int_\Gamma \frac{d\lambda}{2\pi b\lambda} \frac{\psi_{1/\lambda}(x)\psi_{\lambda}(x')}{\epsilon-\epsilon_\lambda}
\end{equation}
Now note that by changing the integration parameter from $\lambda$ to $1/\lambda$ we interchange $x$ and $x'$. Since $\lambda$ and $1/\lambda$ run over the same path $\Gamma$ we can write
\begin{equation}
G_\epsilon(x,x')=\int_\Gamma \frac{d\lambda}{2\pi b\lambda} \frac{\psi_{1/\lambda}(x_<)\psi_{\lambda}(x_>)}{\epsilon-\epsilon_\lambda}.
\end{equation}
We deform now the contour $\Gamma$ towards the origin. Notice that contour goes smoothly over the branch points since $\Gamma$ has components on each pair of Riemann surfaces connected by the branch points. Also, when the contour nears the origin, the integrand goes to zero because $\psi_{1/\lambda}(x_<)\psi_{\lambda}(x>)$ converges to $\lambda^{|x-x'|}$, thanks to the correct ordering of $x$ and $x'$. Thus, the only singularity encountered during the deformation process is when $\epsilon_\lambda$ brushes over $\epsilon$ and from the Residue Theorem we obtain:
\begin{equation}
G_\epsilon(x,x')=\frac{\psi_{1/\lambda}(x_<)\psi_{\lambda}(x_>)}{i b\lambda \partial_\lambda \epsilon_\lambda}.
\end{equation}
If we go back to the $k$ representation, the above expression is the same as the one written in Eq.~\ref{1dg} and this ends our proof for the strictly one dimensional case.

For periodic molecular chains in 3 dimensions, the Riemann surface of the bands was discussed in Ref.~\onlinecite{Prodan:2006yq}, and a typical shape is shown in Fig.~\ref{RiemannS}b. The difference is that now on each disks we can have more than two algebraic branch points and the equation $\epsilon_\lambda = \epsilon$ has an infinite sequence $\epsilon_{\lambda_\alpha}$ of solutions. Starting from the expression
\begin{equation}
G_\epsilon({\bf r},{\bf r})=\int_\Gamma \frac{d\lambda}{2\pi \lambda} \frac{\psi_{1/\lambda}({\bf r}_<)\psi_{\lambda}({\bf r}_>)}{\epsilon_\lambda -\epsilon},
\end{equation}
where $\Gamma$ is the contour shown in Fig.~\ref{RiemannS}b, and  deforming the contour towards the origin and applying the Residue Theorem we obtain:
\begin{equation}
G_\epsilon(x,x')=\sum_\alpha \frac{\psi_{1/\lambda_\alpha}({\bf r}_<)\psi_{\lambda_\alpha}({\bf r}_>)}{i \lambda_\alpha \partial_\lambda \epsilon_{\lambda_\alpha}}.
\end{equation}
In the $k$ representation, this expression becomes:
 \begin{equation}\label{green0}
 G_\epsilon^0(\brr,\brr')=\sum_\alpha\frac{\psi_{-k_\alpha}(\brr_<)\psi_{k_\alpha}(\brr_>)}{i\partial_{k}\epsilon_{k_\alpha}},
 \end{equation}
where $\{k_\alpha\}$ is the infinite sequence of wavenumbers with Im[$k$]$>$0 such that $\epsilon_{k_\alpha}=\epsilon$ and $\brr_<  /\brr_>= \brr / \brr'$ if $z < z'$ and $\brr_<  /\brr_>= \brr'/ \brr$ otherwise.

\subsection{Computing the Green's function for the entire device}
 
We can show in just a few steps why Eq.~\ref{green0} is useful. Indeed, the Green's function for the entire device: $G_\epsilon=(H-\epsilon)^{-1}$ can be computed from the identity:
 \begin{equation}\label{greenf}
 \begin{array}{c}
 G_\epsilon({\bf r},{\bf r}')=G_\epsilon^0({\bf r},{\bf r}')+\int d {\bf r}''\int d{\bf r}''' \medskip \\
 \times G^0({\bf r},{\bf r}'') T_\epsilon({\bf r}'',{\bf r}''') G^0({\bf r}''',{\bf r}'),
\end{array}
 \end{equation}
 where the $T_\epsilon$ matrix is given by
 \begin{equation}\label{Tmatrix}
 T_\epsilon=\Delta V+\Delta V G_\epsilon \Delta V.
 \end{equation}
 Given that $\Delta V$=$\Delta V_L$+$\Delta V_R$, we can naturally decompose the $T$ matrix as
 \begin{equation}
 T_\epsilon=T_{\mbox{\tiny{L}}}+T_{\mbox{\tiny{R}}}+T_{\mbox{\tiny{LR}}}+T_{\mbox{\tiny{RL}}}.
 \end{equation}
Now, the key observation is that, because of the localization properties of $\Delta V_L$ and $\Delta V_R$, by taking $\brr$ and $\brr'$ near the middle of the chain we can tell what is the ordering between ${\bf r}$ and ${\bf r}''$ and between ${\bf r}'''$ and ${\bf r}'$ in the integral of Eq.~\ref{greenf}. Given this, the integrals can be formally executed and the result is:
 \begin{equation}\label{green}
 \begin{array}{c}
 G_\epsilon(\brr,\brr')=G_\epsilon^0(\brr,\brr')+\sum\limits_{\alpha,\beta} \frac{1 }{i\partial_k \epsilon_{k_\alpha} i\partial_k \epsilon_{k_\beta}} \times \medskip \\
 \left \{T_{\mbox{\tiny{L}}}^{\alpha\beta}\psi_{k_\alpha}(\brr)\psi_{k_\beta}(\brr')+T_{\mbox{\tiny{R}}}^{\alpha\beta}\psi_{-k_\alpha}(\brr)\psi_{-k_\beta}(\brr') \right . \medskip \\
\left . +T_{\mbox{\tiny{LR}}}^{\alpha\beta}\psi_{k_\alpha}(\brr)\psi_{-k_\beta}(\brr')+T_{\mbox{\tiny{RL}}}^{\alpha\beta}\psi_{-k_\alpha}(\brr)\psi_{k_\beta}(\brr') \right \}
 \end{array}
\end{equation}
 where
 \begin{equation}
 \begin{array}{c}
  T_{\mbox{\tiny{L}}}^{\alpha \beta} = \int d{\bf r} \int d{\bf r}' \ \psi_{-k_\alpha}({\bf r})T_{\mbox{\tiny{L}}}({\bf r},{\bf r}')\psi_{-k_\beta}({\bf r}')\medskip \\
  T_{\mbox{\tiny{R}}}^{\alpha \beta} = \int d{\bf r} \int d{\bf r}' \ \psi_{k_\alpha}({\bf r})T_{\mbox{\tiny{R}}}({\bf r},{\bf r}')\psi_{k_\beta}({\bf r}')\medskip \\
  T_{\mbox{\tiny{LR}}}^{\alpha \beta} = \int d{\bf r} \int d{\bf r}' \ \psi_{-k_\alpha}({\bf r})T_{\mbox{\tiny{LR}}}({\bf r},{\bf r}')\psi_{k_\beta}({\bf r}')\medskip \\
  T_{\mbox{\tiny{RL}}}^{\alpha \beta} = \int d{\bf r} \int d{\bf r}' \ \psi_{k_\alpha}({\bf r})T_{\mbox{\tiny{RL}}}({\bf r},{\bf r}')\psi_{-k_\beta}({\bf r}')
  \end{array}
 \end{equation}
This is the analytic expression of the Green's function we mentioned at the beginning. The ``$T$''  coefficients remain to be computed numerically, but at this point we have obtained the exact dependence of $G_\epsilon$ on the coordinates $\brr$ and $\brr'$, which will allows us to compute the conductivity tensor. Eq.~\ref{green0} is also essential for deriving the exact asymptotic form of the ``$T$" coefficients in the limit of long chains.\cite{Prodan:2007qv}

Since $H_0$ has no spectrum at $\epsilon_F$, $G_\epsilon^0$ behaves smoothly when $\epsilon$ crosses the real line and consequently (see Eq.~\ref{delta}):
 \begin{equation}\label{green}
 \begin{array}{c}
 \Delta G_{\epsilon_F}(\brr,\brr')=\sum\limits_{\alpha,\beta} \frac{1 }{i\partial_k \epsilon_{k_\alpha} i\partial_k \epsilon_{k_\beta}} \times \medskip \\
 \left \{\Delta T_{\mbox{\tiny{L}}}^{\alpha\beta}\psi_{k_\alpha}(\brr)\psi_{k_\beta}(\brr')+\Delta T_{\mbox{\tiny{R}}}^{\alpha\beta}\psi_{-k_\alpha}(\brr)\psi_{-k_\beta}(\brr') \right . \medskip \\
\left . +\Delta T_{\mbox{\tiny{LR}}}^{\alpha\beta}\psi_{k_\alpha}(\brr)\psi_{-k_\beta}(\brr')+\Delta T_{\mbox{\tiny{RL}}}^{\alpha\beta}\psi_{-k_\alpha}(\brr)\psi_{k_\beta}(\brr') \right \},
 \end{array}
\end{equation}
 where $\Delta T$ stands for $T_{\epsilon_F+i\delta}-T_{\epsilon_F-i\delta}$.
 
 \begin{figure*}
  \includegraphics[width=15cm]{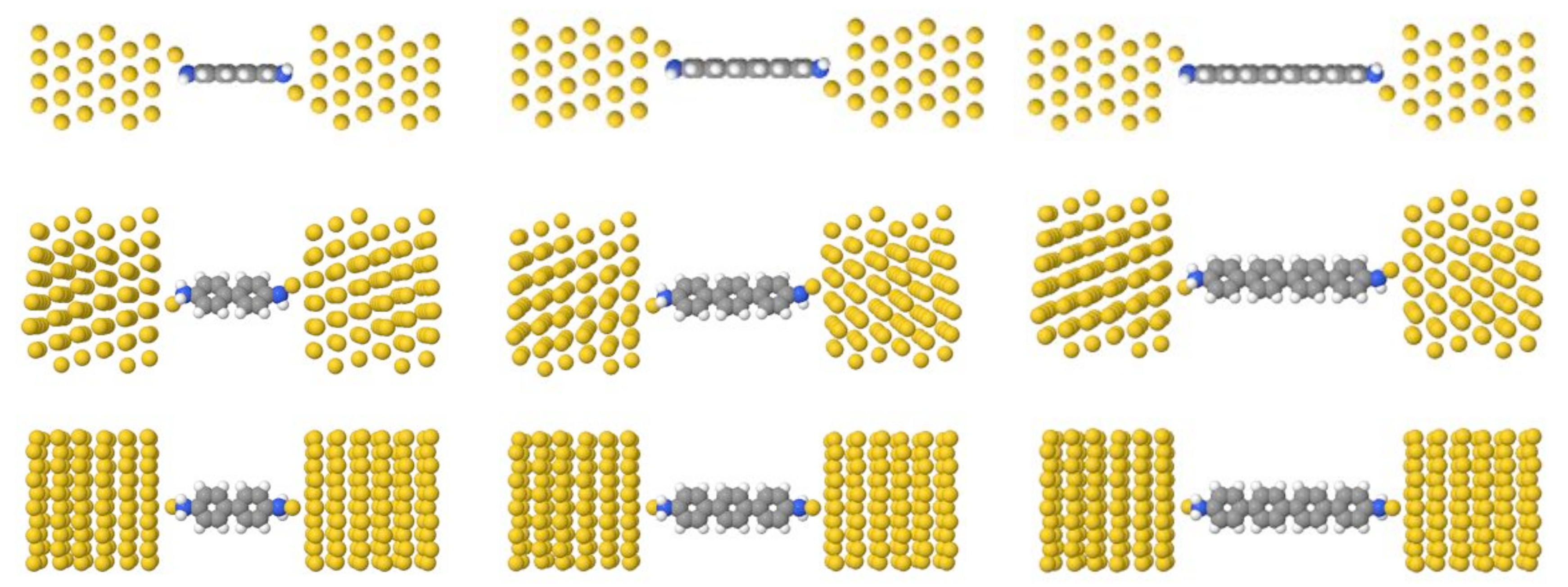}
  \caption{Atomic configurations of the molecular devices. Different rows show different view angles.}
 \label{devices}
\end{figure*}

\subsection{Computing the tunneling conductance}

Given our expression for the Green's function Eq.~\ref{green}, it is evident that the integrals in Eq.~\ref{adiabaticg} lead to generalized Wronskians between different Bloch functions. The generalized Wronskian for two functions $\psi$ and $\phi$ is defined as:
 \begin{equation}
 W(\psi,\phi)=\int d \brr_\bot  \ \psi(\brr_\bot,z)  \overleftrightarrow{\partial_z} \phi(\brr_\bot,z).
 \end{equation}
 We have the following remarkable property,\onlinecite{Prodan:2007qv} valid at arbitrary energy $\epsilon$:
\begin{equation}\label{wrproperty}
\left \{
\begin{array}{l}
W(\psi_{k_\alpha},\psi_{k_\beta})=0 \medskip \\
W(\psi_{k_\alpha},\psi_{-k_\beta})=-\frac{2m}{\hbar^2}i\partial_k \epsilon_{k_\alpha}\delta_{k_\alpha,k_\beta},
\end{array}
\right .
\end{equation}
where $\{k_\alpha\}$ is the sequence of wavenumbers corresponding to the energy $\epsilon$. Applying the above rules, we obtain the following expression for conductance:
\begin{equation}\label{exactg}
G = -\frac{2e^2}{h} \sum_{\alpha,\beta} \frac{\Delta T_{\mbox{\tiny{L}}}^{\alpha\beta} \Delta T_{\mbox{\tiny{R}}}^{\alpha\beta}+\Delta T_{\mbox{\tiny{LR}}}^{\alpha\beta} \Delta T_{\mbox{\tiny{RL}}}^{\alpha\beta} }
 {i\partial_k \epsilon_{k_\alpha} i\partial_k \epsilon_{k_\beta}}
 \end{equation}
The above expression is exact for insulating chains. It does not apply to metallic chains since we used the fact that $H_0$ does not have spectrum at the Fermi level. The matrix elements of $\Delta T$ have simple and intuitive expressions:
\begin{equation}
 \begin{array}{c}
  \Delta T_{\mbox{\tiny{L}}}^{\alpha \beta}=\int d{\bf r} \int d{\bf r}' \medskip \\
  \times \psi_{-k_\alpha}({\bf r}) \Delta V_{\mbox{\tiny{L}}}({\bf r}) \Delta G_{\epsilon_F}({\bf r},{\bf r}')\Delta V_{\mbox{\tiny{L}}}({\bf r}')   \bigskip \\
 \Delta T_{\mbox{\tiny{R}}}^{\alpha \beta}=\int d{\bf r} \int d{\bf r}' \ \psi_{k_\alpha}({\bf r}) \medskip \\
 \times \psi_{-k_\beta} ({\bf r}') \Delta V_{\mbox{\tiny{R}}}({\bf r}) \Delta G_{\epsilon_F}({\bf r},{\bf r}')\Delta V_{\mbox{\tiny{R}}}({\bf r}')  \psi_{k_\beta} ({\bf r}') \bigskip \\
\Delta T_{\mbox{\tiny{LR}}}^{\alpha \beta}=\int d{\bf r} \int d{\bf r}' \medskip \\
\times  \psi_{-k_\alpha}({\bf r}) \Delta V_{\mbox{\tiny{L}}}({\bf r}) \Delta G_{\epsilon_F}({\bf r},{\bf r}')\Delta V_{\mbox{\tiny{R}}}({\bf r}')  \psi_{k_\beta} ({\bf r}')  \bigskip \\
\Delta T_{\mbox{\tiny{RL}}}^{\alpha \beta}=\int d{\bf r} \int d{\bf r}' \medskip \\
\times \psi_{k_\alpha}({\bf r}) \Delta V_{\mbox{\tiny{R}}}({\bf r}) \Delta G_{\epsilon_F}({\bf r},{\bf r}')\Delta V_{\mbox{\tiny{L}}}({\bf r}')  \psi_{-k_\beta} ({\bf r}') 
  \end{array}
 \end{equation}
and they can all be expressed in terms of the spectral operator $\rho_{\epsilon_F}$=$\frac{1}{2\pi i} [ G_{\epsilon_F^+}-G_{\epsilon_F^- }]$. The diagonal $\rho_{\epsilon_F}(\brr,\brr)$ of the spectral operator gives the local density of states. 

As discussed in \onlinecite{Prodan:2007qv}, $\Delta T_{\mbox{\tiny{RL}}}^{\alpha \beta}$ and $\Delta T_{\mbox{\tiny{LR}}}^{\alpha \beta}$ coefficients are exponentially small compared to $\Delta T_{\mbox{\tiny{L}}}^{\alpha \beta}$ and $\Delta T_{\mbox{\tiny{R}}}^{\alpha \beta}$, so in the asymptotic limit of long molecular chains, the tunneling conductance is given by:
\begin{equation}\label{insulatingg0}
G(L) = \frac{2e^2}{h}\sum\limits_{\alpha,\beta} \frac{\Theta_{\mbox{\tiny{L}}}^{\alpha,\beta} \Theta_{\mbox{\tiny{R}}}^{\alpha,\beta}}{i\partial_k \epsilon_{k_\alpha} i\partial_k \epsilon_{k_\beta}}  e^{i(k_\alpha+k_\beta)L},
\end{equation}
with:
\begin{equation}\label{theta1}
\begin{array}{c}
\Theta_{\mbox{\tiny{L}}}^{\alpha \beta}=2\pi  \int d \brr \int d \brr'  \psi_{-k_\alpha}({\bf r}_\bot,z+\frac{L}{2}) \times \medskip \\
\Delta V_{\mbox{\tiny{L}}}(\brr)\rho_{\epsilon_F}(\brr,\brr')\Delta V_{\mbox{\tiny{L}}}(\brr') \psi_{-k_\beta}({\bf r}'_\bot,z'+\frac{L}{2}) \bigskip \\
\Theta_{\mbox{\tiny{R}}}^{\alpha \beta}=2\pi  \int d \brr \int d \brr'  \psi_{k_\alpha}({\bf r}_\bot,z-\frac{L}{2})\times \medskip \\
\Delta V_{\mbox{\tiny{L}}}(\brr)\rho_{\epsilon_F}(\brr,\brr')\Delta V_{\mbox{\tiny{L}}}(\brr') \psi_{k_\beta}({\bf r}'_\bot,z'-\frac{L}{2}).
\end{array}
\end{equation}
In the limit $L\rightarrow \infty$, the $\Theta$ coefficients become independent of $L$. Strictly speaking, the asymptotic form of $G(L)$ is determined by the wavenumber $k$ with  minimum imaginary component. This is the case for the phenyl chains that we will investigate in the next Section, or for alkyl chains that were investigated in Ref.~\onlinecite{Prodan:2008by}. However, for more complex molecular chains such as carbon nanotubes \onlinecite{Pomorski:2004nx}, there may be many wavenumbers with similar imaginary parts, in which case we must consider more than one evanescent channel in Eq.~\ref{insulatingg0}. We point out that Eq.~\ref{insulatingg0} tells how the evanescent tunnels interfere with each other during tunneling transport.

It is important to observe that computing the contact conductance requires a converged density of states near the contacts, which can be obtained from a standard supercell calculation that includes large enough electrodes. The spectral operator $\rho_{\epsilon_F}(\brr,\brr')$ can be computed in various ways and each way can have its advantages and disadvantages. Provided one can store a large number of orbitals, a straightforward way consists in using the Kohn-Sham orbitals $\phi_\epsilon$:
\begin{equation}\label{spectral}
\rho_{\epsilon_F}(\brr,\brr')=\frac{1}{\pi}\sum\limits_\epsilon \frac{ \delta}{(\epsilon-\epsilon_F)^2+\delta^2} \phi_\epsilon^*({\bf r}) \phi_\epsilon({\bf r}'),
\end{equation}
which leads to:
\begin{equation}\label{finalt1}
\begin{array}{c}
\Theta_{\mbox{\tiny{L}}}^{\alpha \beta}=  \sum\limits_\epsilon \frac{2 \delta}{(\epsilon-\epsilon_F)^2+\delta^2} \int d \brr 
 \ \psi_{-k_\alpha}({\bf r}_\bot,z+\frac{L}{2})\Delta V_{\mbox{\tiny{L}}}(\brr) \phi_\epsilon^*({\bf r})  \medskip \\
 \times \int d \brr  \ \phi_\epsilon(\brr) \Delta V_{\mbox{\tiny{L}}}(\brr)\psi_{-k_\beta}({\bf r}_\bot,z+\frac{L}{2})
\end{array}
\end{equation}
and
\begin{equation}\label{finalt2}
\begin{array}{c}
\Theta_{\mbox{\tiny{R}}}^{\alpha \beta}=  \sum\limits_\epsilon \frac{2 \delta}{(\epsilon-\epsilon_F)^2+\delta^2} \int d \brr  
 \ \psi_{k_\alpha}({\bf r}_\bot,z-\frac{L}{2})\Delta V_{\mbox{\tiny{R}}}(\brr) \phi^*_\epsilon({\bf r}) \medskip \\
 \times\int d \brr \ \phi_\epsilon(\brr) \Delta V_{\mbox{\tiny{R}}}(\brr)\psi_{k_\beta}({\bf r}_\bot,z-\frac{L}{2}).
\end{array}
\end{equation}
This is the way we actually compute the coefficients in this work and details about how we choose $\delta$ will be given later in the paper. An alternative way will be to compute the spectral operator directly from the Green's functions. This involves inverting the large matrices $(H-\epsilon \pm \delta)^{-1}$, which can be done iteratively and would not require saving large amounts of data.

\section{Application to devices involving phenyl chains} 

In the following, we present an application to devices made of phenyl chains attached to gold electrodes via amine groups, like the those investigated in Ref.~\onlinecite{Venkataraman:2006db}.  The complex band structure calculations of Ref.~\onlinecite{Fagas:2004ao} reveal an evanescent channel with Im[$k$] much smaller than that of the rest of the channels. Consequently, the tunneling conductance is determined by this evanescent channel and the expression for the tunneling conductance simplifies to:
\begin{equation}\label{Gphenyl}
G = \Theta_{\mbox{\tiny{L}}} \Theta_{\mbox{\tiny{R}}}  e^{2ikL}.
\end{equation}
This is to be compared to the classical expression $G$=$G_ce^{-\beta N}$. The tunneling coefficient $\beta$ is related to $k$ via $\beta=2\mbox{Im}[k]b$. The contact conductance $G_c$ is given by the pre-exponential factor in Eq.~\ref{Gphenyl}. To be precise, let us write the simplified expression of theta coefficients
\begin{equation}\label{theta1}
\begin{array}{c}
\Theta_{\mbox{\tiny{L}}}=\frac{2\pi}{W(\psi_k,\psi_{-k})} \int d \brr \int d \brr'  \times \medskip \\
\psi_{-k}(\brr)\Delta V_{\mbox{\tiny{L}}}(\brr)\rho_{\epsilon_F}(\brr,\brr')\Delta V_{\mbox{\tiny{L}}}(\brr')\psi_{-k}(\brr'),
\end{array}
\end{equation}
with $\brr$ and $\brr'$ measured from the left end of the chain. Similarly
\begin{equation}\label{theta2}
\begin{array}{c}
\Theta_{\mbox{\tiny{R}}}=\frac{2\pi}{W(\psi_k,\psi_{-k})} \int d \brr \int d \brr'  \times \medskip \\
\psi_{k}(\brr)\Delta V_{\mbox{\tiny{R}}}(\brr)\rho_{\epsilon_F}(\brr,\brr')\Delta V_{\mbox{\tiny{R}}}(\brr')\psi_{k}(\brr'),
\end{array}
\end{equation}
with $\brr$ and $\brr'$ measured from the right end of the chain. We have included the derivatives $i\partial_k \epsilon_k$ into the $\Theta$ coefficients, and then we expressed this derivatives using the generalized Wronskian. This particular way of writing the $\Theta$ coefficients is useful since the formulas become independent of the normalization of the evanescent waves. 

\begin{figure}
\center
  \includegraphics[width=8.6cm]{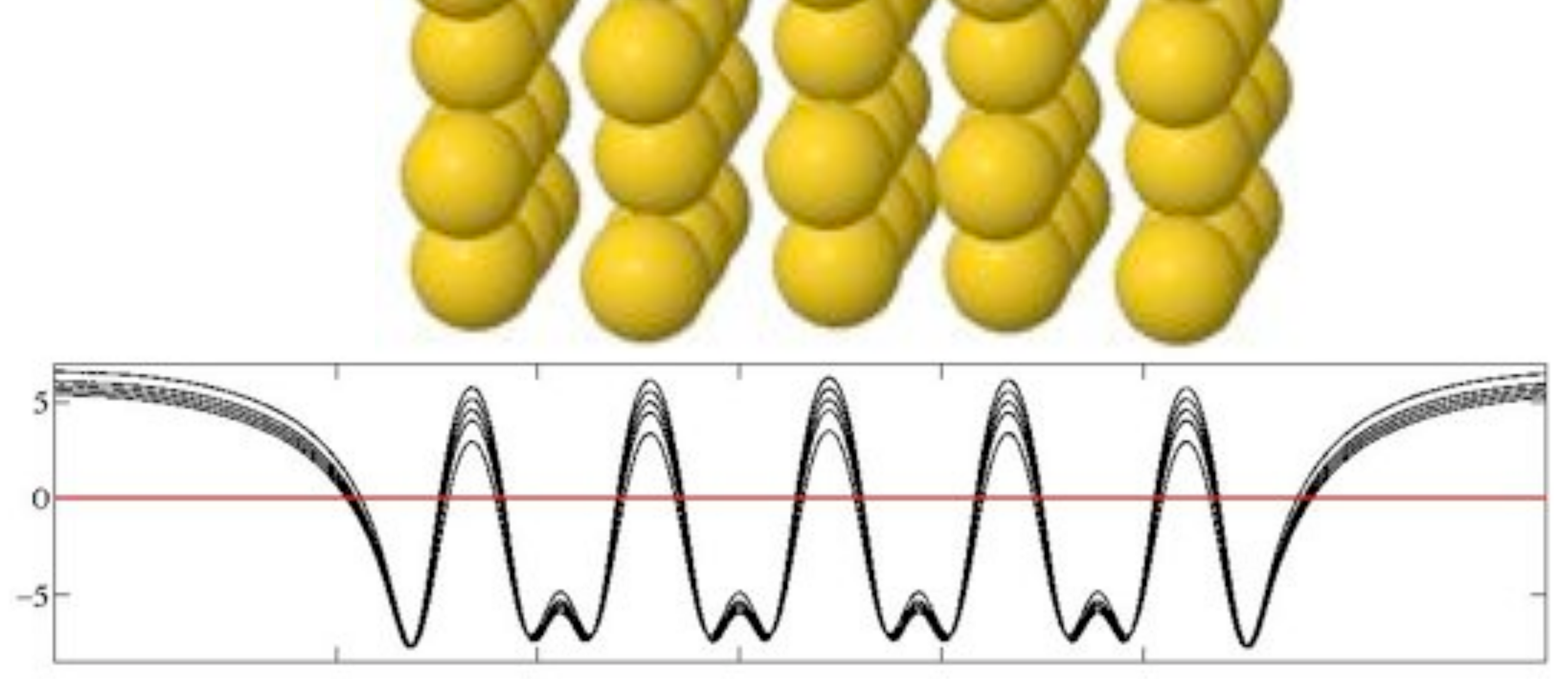}\\
  \caption{A plot of the lateral average of the effective potential of a 5 layer Au slab, corresponding to various values of parameter $\alpha$. The vacuum region around the slab is much larger than what is visible in the picture.}
 \label{CoreCorrection}
\end{figure}

\subsection{Computational details} 

We study three devices, containing 2, 3 and 4 phenyl rings, linked to gold electrodes via amine groups. These three devices will be referred to as (a), (b) and (c), respectively. The corresponding atomic configurations are shown in Fig.~\ref{devices}. Only the planar configuration for the phenyl chain will be considered. The geometry of the planar phenyl chain was build from the structure of biphenyl molecule reported in Ref.~\onlinecite{Stand.:lr}. This reference reports an average C-C bond length of 1.40 \AA  \ for the ring C atoms and a separation between the phenyl rings of 1.49 \AA. With this bond lengths, the unit cell size of the chain is  4.315 \AA \ in $z$ direction. The C-H  bond length was fixed at 1.10 \AA. The bond lengths reported in Ref.~\onlinecite{Stand.:lr} are weakly dependent on the functional and basis set being used in the calculations.

The bond angles for the N atoms of the linking groups were fixed in a tetrahedral configuration, except for the bond with the Au atom. The N-C and N-H bond lengths were fixed at 1.41 \AA \ and 1.04 \AA, respectively. The Au-N bond length was fixed at 2.40 \AA \ and the C-N-Au bond angle was fixed at 123$^o$. Indicating by A, B, and C the stacking planes in the (111) direction for fcc Au, the devices can be represented schematically by:
\begin{equation}
\mbox{CBACBA-Au-NH$_2$-(C$_6$H$_4$)$_N$-NH$_2$-Au-CBACBA}
\end{equation}
Ideally, the left (right) Au ad-atom would occupy a lattice site of the C (A) stacking plane. Because of computational constrains that require the chain to be oriented along the $z$ direction and the surface of the electrode to be perpendicular to the $z$ direction, this ideal configuration cannot be exactly satisfied, instead the ad-atoms are displaced towards the chain's plane by about 0.5 \AA. Since we are interested here mainly in illustrating the method, we did not investigate the issue of how geometrical factors such as the accurate position of the adatoms and distortions of the phenyl chain affect the calculated conductance. These issues are, however, very important for accurate quantitative comparisons with experiment. 
   
The lattice constant for the gold atoms in the leads was fixed at the experimental value (thus the stacking planes are spaced by 2.35 \AA). No surface reconstruction was considered. The system is periodically repeated in all three direction, but the calculations are restricted to the $\Gamma$ point. In the $z$ direction, the periodically repeated system has 12 layers of Au between two consecutive phenyl chains. For such electrode size, we expect the density of states near the contacts to be well converged. The lateral size of the supercell was chosen so that 20 Au atoms are contained in each layer. Thus, our computational supercell contains 242 Au atoms. In total, there are 268, 278 and 288 atoms for devices (a), (b) and (c), respectively.  

The equilibrium self-consistent Kohn-Sham calculations were performed with a real space, pseudopotential code based on finite differences. The same code was used for the calculations reported in Ref.~\onlinecite{Prodan:2008by}. We adopted a 5-point finite difference approximation for the kinetic energy operator, and used a uniform rectangular space grid with a spacing of 0.3547 a.u., sufficient for a good convergence of the electronic structure. This grid is commensurate with the unit cell of the periodic phenyl chain, which is the reference system in our transport calculations. We adopted the Local Density Approximation (LDA) for exchange and correlation using the Perdew-Zunger (PZ)\cite{Perdew:1981gb} interpolation of the numerical electron-gas data of Ceperley and Alder.\cite{Ceperley:1980eu} We used Troullier-Martin norm-conserving pseudo-potentials\cite{Troullier:1991ys} for all the atomic species. The pseudopotentials for C and N atoms had distinct {\it s} and {\it p} components and we took the {\it p} pseudo-potential as the local reference. Purely local pseudopotentials were used for the H and Au atoms. In the latter case only the outermost {\it s} electrons were treated explicitly.

\begin{figure*}
\center
  \includegraphics[width=15cm]{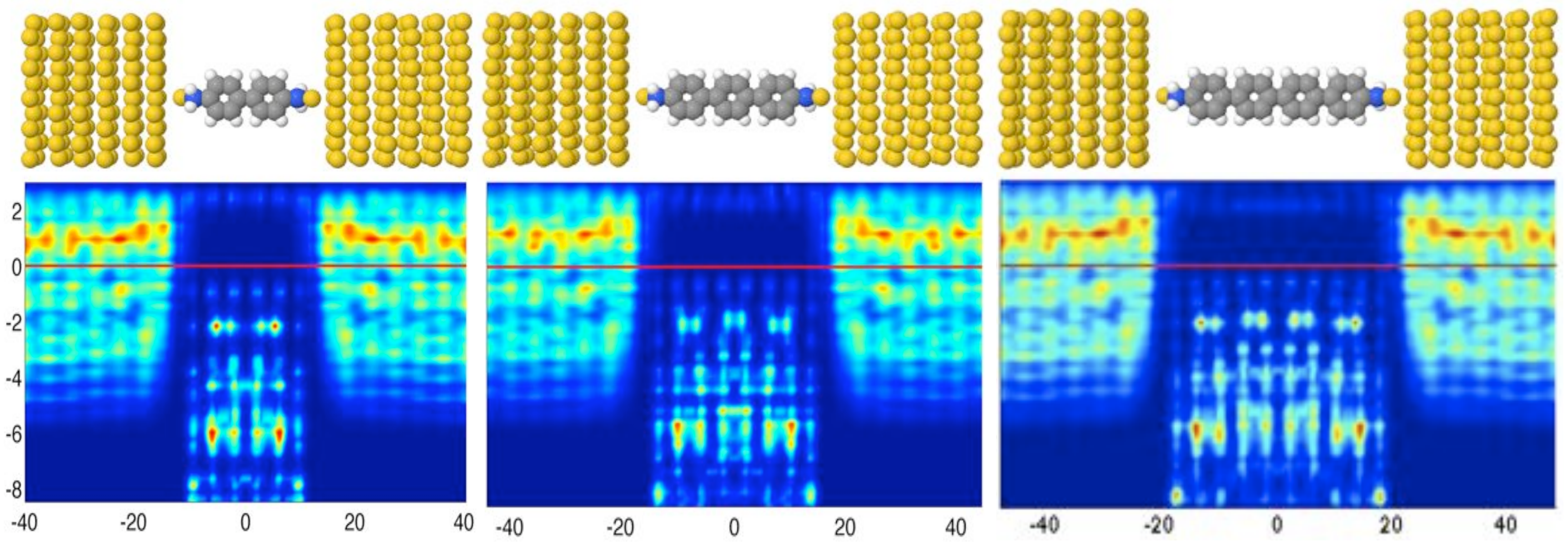}\\
  \caption{The planar average (over xy) of the local density of states, shown as a density plot with energy (in eV) on the vertical axis and z coordinate (in a.u.) on the horizontal axis.}
 \label{NewLdos}
\end{figure*}

Since the current calculations include only the {\it s} electrons of Au, the calculated work function of the leads differs from the experimental value. To address this problem, non-linear core corrections were proposed in Ref.~\onlinecite{Louie:1982hl}. Even with these corrections, the work function of fcc Au, as given by LDA calculations, takes values between 6.28 and 6.70 eV, depending on the surface orientation \onlinecite{Fall:2000kx}. On the other hand, when the {\it d} Au electrons are treated explicitly, the LDA yields\cite{Fall:2000kx} workfunctions close to experiment.\cite{Hansson:1978ij} In our calculation, with the non-linear core corrections implemented as in Ref.~\onlinecite{Louie:1982hl}, we find a work function of 6.6 eV for the Au electrodes, which should be compared to an average experimental value of $~5.4$ eV for the workfunction of the Au (111) surface.\cite{Hansson:1978ij} While this difference had insignificant consequences for the alkyl chains,\cite{Prodan:2008by} due to their large insulating gap and to the particularities of their complex band structure, for phenyl chains the consequences will be more severe due to their smaller insulating gap and to the parabolic shape of the complex band. More precisely, the Fermi level of the device will be located extremely close to the edge of the valence band of the insulating chain.

Since our main purpose here is to demonstrate our methodology, we adopted a simple empirical approach to correct this shortcoming: we modified the local pseudo-potential of Au atoms by adding a local core correction of the form $\alpha n_d({\bf r})$ (Ry), where $n_d({\bf r})$ is the density of the frozen Au $d$ electrons. The work function for the Au (111) surface becomes 5.4 eV if the constant $\alpha$ is fixed at 2.5 Ry$\times$Bohr$^3$. Fig.~\ref{CoreCorrection} shows plots of the effective potential of a 5 layer Au slab for increasing values of $\alpha$. Here we can see a monotonic bending of the potential in the vacuum region, leading to a reduction of the workfunction. We can also see a relatively large change inside the $d$ cores, but these changes have minor effects on the occupied electron density since they occur well above the Fermi level. In addition, we do see a small change in the potential in between the planes.

\subsection{Electronic Structure}

The results of the electronic structure calculations are summarized in Fig.~\ref{NewLdos}, which illustrates the local density of states for the three devices, averaged in the xy plane: $\rho_{\mbox{av}}(z,\epsilon)$=$\int \rho_\epsilon(x,y,z)dxdy$. The plots give a color map of $\rho_{\mbox{av}}(z,\epsilon)$ in the plane of energy $\epsilon$ and of position $z$. The figure was constructed from all Kohn-Sham orbitals used in the transport calculations, their number being equal to the number of the occupied orbitals plus additional 110 un-occupied orbitals (without counting the spin). The Fermi level was fixed at zero and is indicated by the red line. In these plots, the conducting states of the leads and the band edges of the insulating chain are quite visible. The Fermi level, which is pinned by the continuum states of the leads, falls into the insulating gap of the phenyl chain. One sees that the conducting states of the leads decay rapidly to zero inside the phenyl chain, where the spectral gap becomes visible. The gap is clean all the way to the first gold atoms of the electrodes, showing no surface resonances. For energies inside the spectral gap of the chain, $\rho_{\mbox{av}}(z,\epsilon)$ does not show any special features near the contacts. The insulating band gap seen in Fig.~\ref{NewLdos} is larger for device (a) and is comparable for devices (b) and (c). The tunneling transport is sensitive to the Fermi level alignment relative to the edges of the insulating gap. We point the reader to the Refs. \onlinecite{McDermott:2009pi,Wang:2008eu}, which give an extended discussion of the band alignment in molecular electronic devices and its effect on transport.

\begin{figure}
\center
  \includegraphics[width=8.6cm]{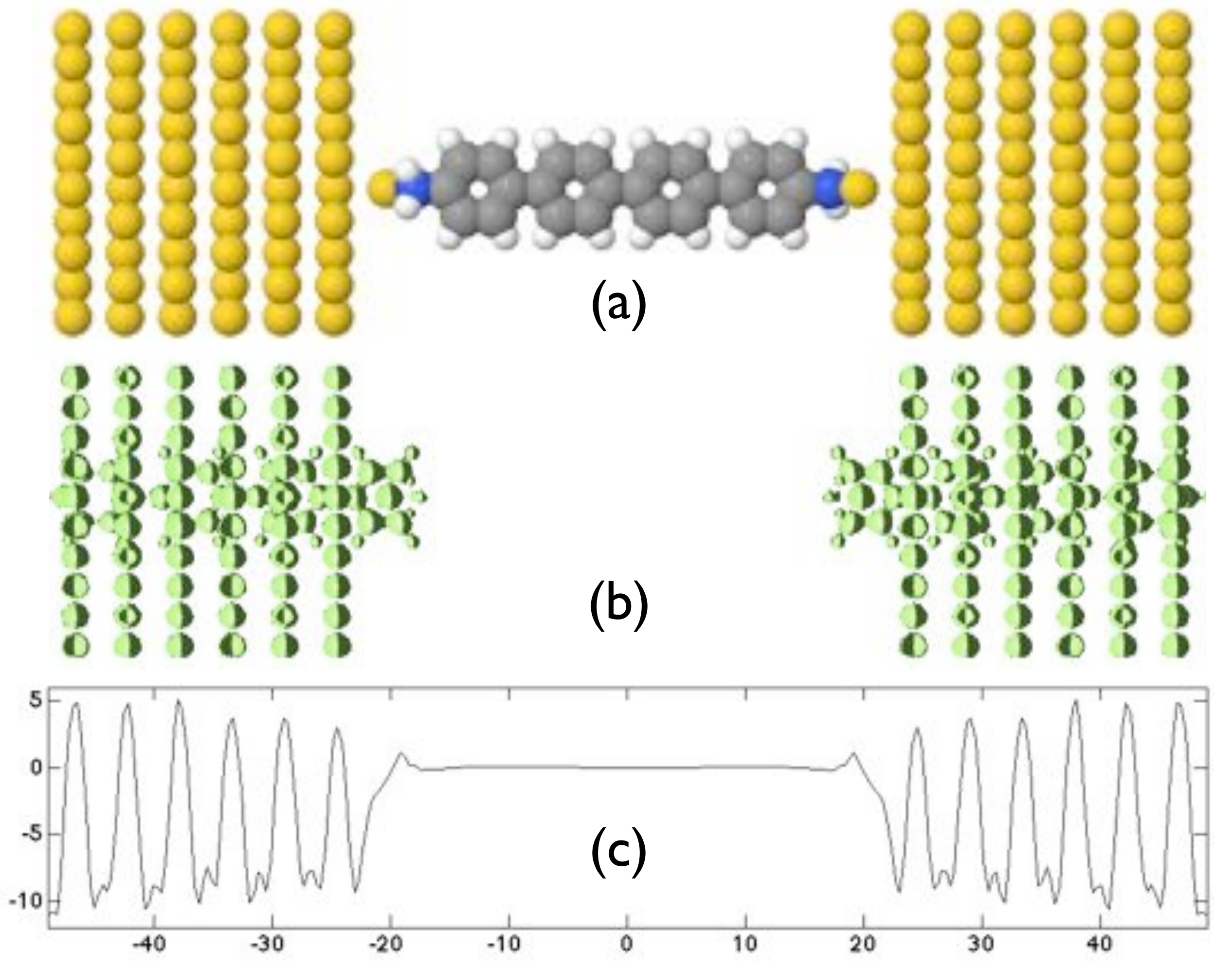}\\
  \caption{a) Atomic configuration of device (c). b) An iso-surface of $V_{\mbox{eff}}$. c) Planar average of $V_{\mbox{eff}}$ (with respect to the xy coordinates). d) An iso-surface of $\Delta V$. e) Planar average of $\Delta V$ (with respect to the xy coordinates). The energy units are Ry.}
 \label{DiffPot}
\end{figure}

Fig.~\ref{DiffPot} illustrates the local part of the $\Delta V$, confirming the main assumption behind our formalism, namely that the potential inside the insulating chain is, to a very high degree, periodic and that $\Delta V$ is localized on the leads. Since we use norm conserving pseudo-potentials, the non-local part of $\Delta V$ is automatically localized on the leads.

The band structure of the periodic potential $V_0$ for device (c) is shown in Fig.~\ref{bands}. The real and complex structures are similar to those reported in Ref.~\onlinecite{Fagas:2004ao}, at least for energies below the vacuum. Above the vacuum, our calculation shows additional bands originating from scattering states, which are absent in the tight binding calculations of Ref.~\onlinecite{Fagas:2004ao}.

\subsection{The conductance: Numerical results.}

We would like to comment first on the numerical advantages brought in by our formalism. Due to the large supercells involved in this kind of calculations, very often transport calculations for long molecular chains are carried out in a reduced basis set representation of the Hilbert space of the electron states. This can be problematic because the basis set functions are usually localized and it is not always clear how well are the scattering states represented by a small number of such functions. When considering experimental values for $G$ that are between $10^{-3}$ and $10^{-6} G_0$ or even smaller, one can easily see that there is very little margin for errors. In our calculations, all the quantities involved in the formula for $G$ are computed on the same grid used for the self-consistent calculation. Since the asymptotic expression of $G$ is virtually exact for long chains, the analytic formula of Eq.~\ref{insulatingg0} allows us to compute $G$ without truncating our Hilbert space. 

\begin{figure}
 \center
  \includegraphics[width=8.6cm]{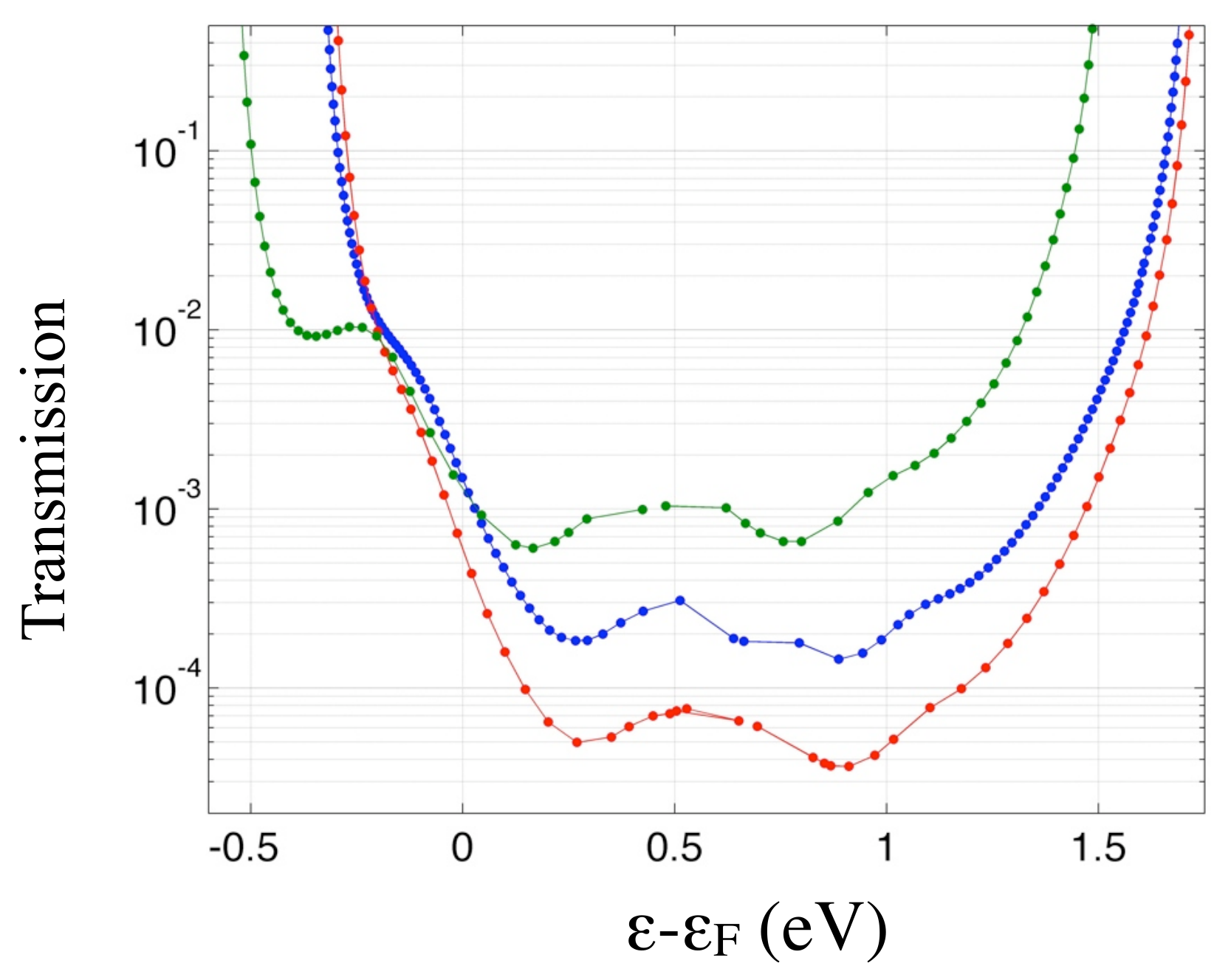}\\
  \caption{Plots of the transmission as function of energy; green, blue and red colors are used for devices (a), (b) and (c), respectively.}
 \label{gVn}
\end{figure}

We computed the transmission coefficient of our devices by evaluating Eq.~\ref{Gphenyl} at  several energies $\epsilon$ within the insulating gap and the results are reported in Fig.~\ref{gVn} as a function of $\epsilon$-$\epsilon_F$. We should point out that the computed values become less accurate for energies closer to the band edges. The calculated transmission of the device (a) looks different from the others, mainly because of its larger insulating gap. The linear conductance of the three devices, as derived from these calculations, are $G=$ $1.5\times 10^{-3}$, $1.5\times 10^{-3}$, $4.3 \times 10^{-4}G_0$, respectively. The $\beta$ coefficient, computed as 2$b$Im[${k_F}$], is equal to 1.15 for device (a) and 0.98 for the other two devices. It appears that only the last two devices reached the asymptotic tunneling regime. However, the situation is highly dependent on the position of the Fermi level. For example, $\beta$ would be the same for the three devices if the Fermi level would move away from the valence band edge of the phenyl chain by 0.2 eV. Since the values of $G$ are highly sensitive to the band alignment, we should be cautious when comparing the theoretical predictions with the experimental values.  In any case, the predicted $G$ for device (a) is very close to the value measured in Ref.~\onlinecite{Venkataraman:2006db}. The predicted value of device (b) is 8.3 times larger than the experimental value reported in  Ref.~\onlinecite{Venkataraman:2006db}. No experimental value has been reported for the device (c). It is interesting to remark that a previous study\cite{Quek:2007xq} on a device consisting of a single phenyl molecule linked to gold electrodes via amine groups predicted a theoretical $G$ that is 7 times larger than the measured experimental value. The same reference pointed out that the calculated DFT conductance would become comparable to the experimental value if the Fermi level were located 0.5 eV further away from the valence band. We also see from our data that a shift of $\epsilon_F$ by 0.5 eV  would bring the theoretical prediction for both devices (a) and (b) in line with the experimental values.
 
\begin{figure}
\center
  \includegraphics[width=6cm]{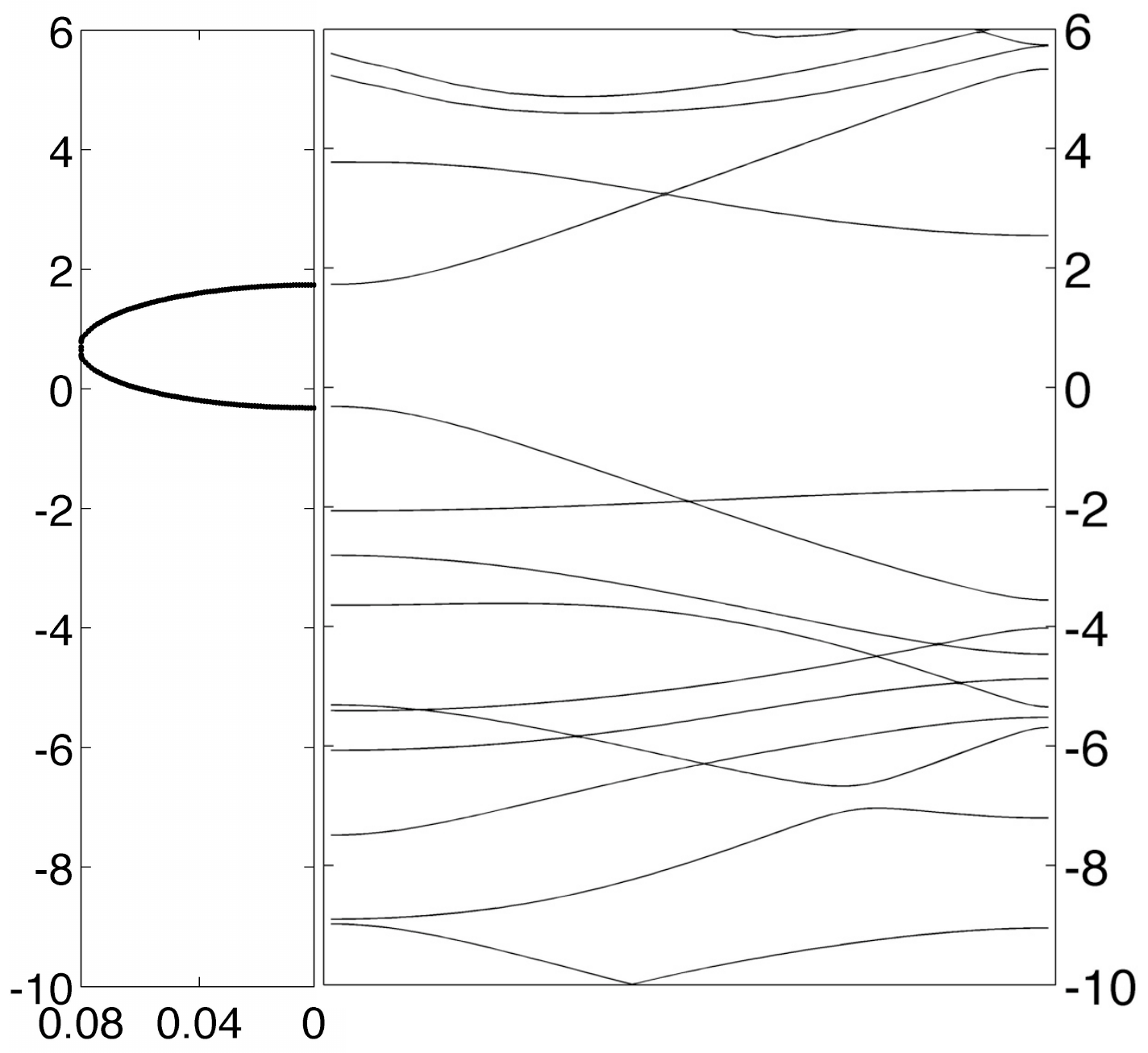}\\
  \caption{Real (right panel) and complex (left panel) band structures corresponding to the periodic potential $V_0$ for device (d). Only the complex band with smallest Im[$k$] is shown. The Fermi level of the device was set to zero. The energy unit is eV and the unit for Im[$k$] is 1/Bohr.}
 \label{bands}
\end{figure}

We now describe how we computed the conductance. The complex band structure corresponding to $V_0$ varies slightly when different devices are considered. Overall, the band structure for $V_0$ is similar to that reported in Ref.~\onlinecite{Fagas:2004ao} for the infinite, isolated phenyl chains, suggesting that the main difference between $V_0$ and  the effective potential of the infinite, isolated chain is a rigid shift. Given the particular complex band structure of the phenyl chains, the tunneling conductance is determined by just one complex band, the one with the smallest Im[$k$]. This complex band is shown in Fig.~\ref{bands} for device (c). It was obtained by varying continuously Im[$k$] from 0 to its maximum value, while keeping Re[$k$]$=0$. For each complex value of $k$, the spectrum of the $k$ dependent Hamiltonian:
\begin{equation}
H_k=-(\nabla-ik{\bf e}_z)^2+V_0+e^{-ik(z-z')}V_{non-loc}({\bf r},{\bf r}'),
\end{equation}
with periodic boundary conditions at $z=\pm b/2$, was calculated and its eigenvalues ordered according to their real parts: Re[$\epsilon_{1k}$]$<$Re[$\epsilon_{2k}$]$<$ \ldots . We focus, in particular, on the 14th  and 15th eigenvalues $\epsilon_{14k}$ and $\epsilon_{15k}$ (which take real values, see Fig.~\ref{bands}) and their corresponding evanescent Bloch functions $\psi_{14k}$ and $\psi_{15k}$. When Im[$k$]=0, $\epsilon_{14k}$ and $\epsilon_{15k}$ coincide, respectively, with the top of the valence bands and with the bottom of the conduction bands of $V_0$.  By increasing Im[$k$], the two eigenvalues move towards each other until they become degenerate when $k$ reaches the branch point at Im[$k$]=0.08 Bohr$^{-1}$. At different values of Im[$k$], we evaluated Eq.~\ref{insulatingg0} for both $\epsilon$=$\epsilon_{14k}$ and $\epsilon$=$\epsilon_{15k}$, using the corresponding evanescent Bloch functions $\psi_{14k}$ and $\psi_{15k}$ to compute the $\Theta$ coefficients via formulas \ref{theta1} and \ref{theta2}. The spectral kernel was computed directly from the Kohn-Sham orbitals of the full device as previously explained. The coefficient $\delta$ was fixed at 0.01 Ry. This value is about an order of magnitude larger than the average energy level spacing of the Kohn-Sham orbitals near the Fermi energy.

\subsection{Insight into the transport properties of phenyl chains}

The analytic result of Eqs.~\ref{insulatingg0}, \ref{theta1} and \ref{theta2} allows us to point several key aspects of the tunneling transport of our devices. Since the formulas involve overlap integrals,  the new insight is obtained by looking at each physical quantity entering in the expressions of the $\Theta$ coeffiencients. 

\begin{figure}
\center
  \includegraphics[width=8.6cm]{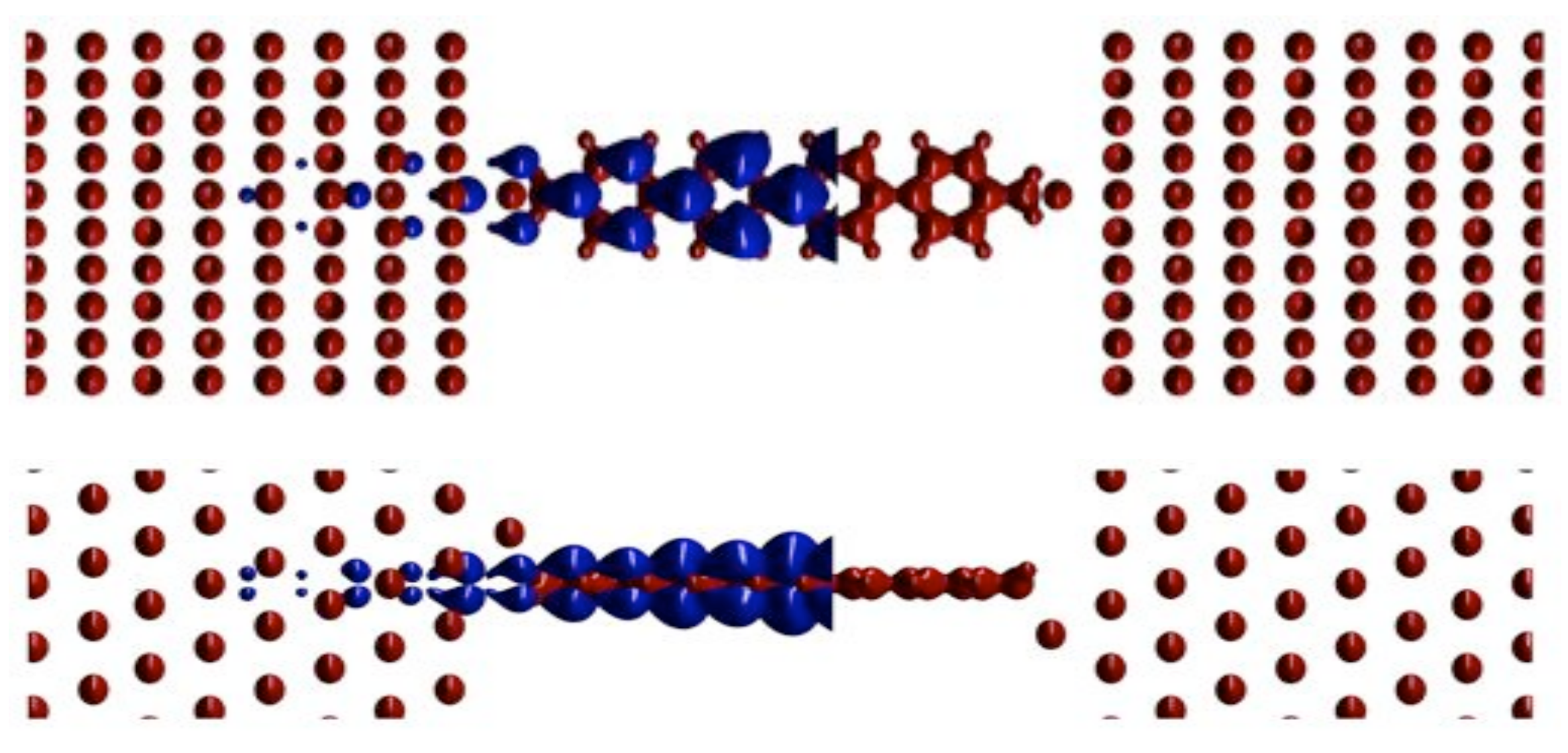}\\
  \caption{Different angle views of an iso-surface plot (blue surface) corresponding to 5\% of the maximum value of the evanescent Bloch function $|\psi_{- k}({\bf r})|$ at the Fermi energy for device (c). For illustrative convenience, the evanescent Bloch function $|\psi_{- k}({\bf r})|$ was truncated at the right end. For reference, we also show an iso-surface plot of the effective potential (in red). The iso-surface plots for $|\psi_{k}({\bf r})|$ can be obtained by mirror symmetry relative to the center of the device.}
 \label{bloch}
\end{figure}

A plot of the local density of states (i.e. the diagonal part of the spectral operator) was already given in Fig.~\ref{NewLdos} and a plot of $|\Delta V|$ was given in Fig.~\ref{DiffPot}. Fig.~\ref{bloch} shows a plot of the evanescent Bloch solutions of the periodic Hamiltonian with potential $V_0$ for device (c), evaluated at the Fermi level. These functions are a property of the periodic Hamiltonian only, but their spatial decay is fixed by the $\beta$ coefficient, which depends on the level alignment as discussed earlier. The contact conductance $G_c$ depends on the overlap of these evanescent functions with other physical quantities, and a plot like the one in Fig.~\ref{bloch} allows us to assess quantitatively the contact region that is relevant to tunneling transport.   

\begin{figure}
\center
  \includegraphics[width=8.6cm]{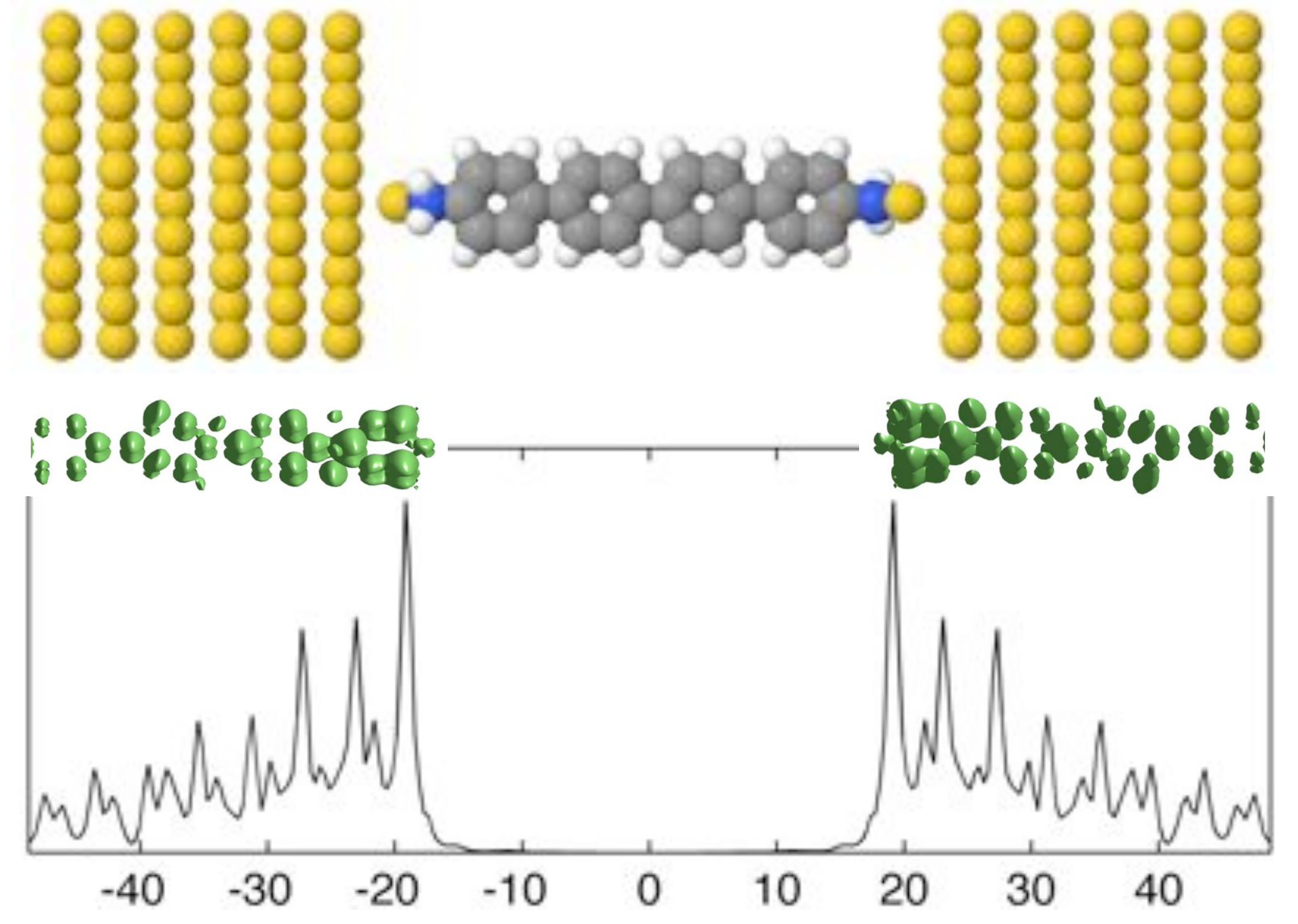}\\
  \caption{An iso-surface plot of $|\Psi_{\mbox{\tiny{L/R}}}({\bf r})|$, corresponding to 2\% of the maximum value of $|\Psi_{\mbox{\tiny{L/R}}}({\bf r})|$ and the planar average of $|\Psi_{\mbox{\tiny{L/R}}}({\bf r})|$ (with respect to the xy coordinates). The horizontal axes of the graphs are aligned. For convenience, we also included the atomic configuration of the device.}
 \label{Psi}
\end{figure}

A main factor in our transport calculation is the overlap between the evanescent Bloch function $\psi_{\mp k}({\bf r})$ and $\Delta V_{\mbox{\tiny{L/R}}}$, i.e. the quantity  
\begin{equation}
\Psi_{\mbox{\tiny{L/R}}}({\bf r}) = \psi_{\mp k}(\brr)\Delta V_{\mbox{\tiny{L/R}}}(\brr),
\end{equation}
which is exponentially localized at the left/right contacts. As a consequence the spectral operator in Eq.~\ref{spectral} is only needed in a region near the contacts. A plot of $\Psi_{\mbox{\tiny{L/R}}}({\bf r})$ for device (c) is shown in Fig.~\ref{Psi}. This plot allows us to understand how the different Au layers contribute to the contact conductance $G_c$.\cite{Prodan:2008by} From the data we extract that the contact Au atom and the next two gold layers contribute to $\Psi_{\mbox{\tiny{L/R}}}({\bf r})$ by about 75\%, while the remaining 25\% comes from the remaining layers. This information tells us that the conductance of our devices is primarily determined by the first three layers of Au atoms, an information that could be useful when designing molecular circuits based on phenyls. The spatial spread of $\Psi_{\mbox{\tiny{L/R}}}({\bf r})$ along the device seen in Fig.~\ref{Psi} is more extended than the one found for alkyl based devices. This implies that the conductance of the present devices is more sensitive to the geometrical and chemical configuration of the contact, or to the orientation of the molecule relative to the molecular wires.

\section{Conclusions} 
 
In conclusion, we presented an extended discussion of a previously proposed theoretical approach for off-resonant tunneling transport. We added details where necessary and we greatly simplified the derivation of the asymptotic expression for the tunneling conductance. In addition, we reported a formally exact expression for the linear conductance derived within the TDCDFT.  

The application to devices involving phenyl chains revealed several interesting facts. Based on our data, it appears that only devices (b) and (c) reached the tunneling regime. Experimentally, one observes that the conductance of devices containing one, two and three phenyls obey the tunneling conductance formula $G$=$G_ce^{-\beta N}$ with same $G_c$ and $\beta$. This can be a coincidence or it can be the real fact. Only measurements on longer phenyl chains can clarify the point.  

We found that the transport calculations are extremely sensitive to band alignment. This is prompted in the first place by the relatively small insulating gap of the phenyl chain but also by the fact that LDA places the Fermi level close to the edge of the valence band of the phenyl chain. For this reason, $k_F$ is located in the rapidly varying region of the complex band and small variations in $\epsilon_F$ lead to large variations in conductance.

The analytic expression for the tunneling conductance allowed us to probe several transport characteristics of the devices. We showed that the contact conductance is exponentially localized near the contacts and we were able to describe quantitatively this localization. Since the evanescent conducting channels decay slower  than for the case of alkyl devices, the contact conductance is less localized and the tunneling characteristics of the phenyl based devices are predicted to be more sensitive to the particularities of the electrodes when compared to devices involving alkyl chains.

\medskip

\noindent{\it Acknowledgments:} Partial support for this work was provided by the 
NSF-MRSEC program through the Princeton Center for 
Complex Materials (PCCM), grant DMR 0213706, and 
by DOE through grant DE-FG02-05ER46201. E.P acknowledges an award from Research Corporation for Science Advancement.

\medskip


\end{document}